\documentclass{ws-ijmpd}
\usepackage{amssymb,graphicx,amsmath,graphics,xfrac,url,subfigure,booktabs,adjustbox}
\usepackage[super,compress]{cite}

\newcommand\sr{\mathrm{sr}}

\newcommand\sca{\mathrm{S}}
\newcommand\ten{\mathrm{T}}

\newcommand\D{\mathrm{d}}

\newcommand\ex{\mathrm{ex}}

\begin{document}
	
\title{Numerical analysis of the generalized Starobinsky inflationary model}
	
\author{Stefano Meza}
\address{Yachay Tech University, School of Physical Sciences and Nanotechnology, Hda. San Jos\'e s/n y Proyecto Yachay, 100119, Urcuqu\'i, Ecuador}

\author{David Altamirano}
\address{Yachay Tech University, School of Physical Sciences and Nanotechnology, Hda. San Jos\'e s/n y Proyecto Yachay, 100119, Urcuqu\'i, Ecuador}

\author{Muhammad Zahid Mughal}
\address{University of GUJRAT, Department of mathematics,  Jalalpur Jattan Road Gujrat, Pakistan}

\author{Clara Rojas \thanks{crojas@yachaytech.edu.ec}}
\address{Yachay Tech University, School of Physical Sciences and Nanotechnology, Hda. San Jos\'e s/n y Proyecto Yachay, 100119, Urcuqu\'i, Ecuador}
	
\maketitle
	
\begin{history}
\received{\today}
\end{history}
	
\begin{abstract}
In this work we study numerically  one kind of generalization of the Starobinsky inflationary model (power-law type), which is characterized by the parameter $p$.  In order to find the parameter $p$ that fixes with observations, we compute the cosmological parameters $A_S$, $n_S$, and $r$ for several values of $p\simeq 1$. We have found that the value of $p=1.0004$ reproduces the value of $A_S$, $n_\sca$, and $r$ in agreement with current observational data.
		
\end{abstract}
	
\maketitle

\maketitle

%*****************
\section{Introduction}
%*****************

Inflation is a stage in the evolution of the Universe where the expansion was accelerated  \cite{liddle:2000}. It was introduced by Alan Guth in the  eighties  \cite{guth:1981}  with the original motivation of solving the problems of the big-bang theory, like the flatness problem, the  horizon problem, and the problem of magnetic monopoles \cite{martin:2019}. Soon was discovered that inflation has another important property,   this theory predicts  curvature perturbations with an almost  scale invariant power spectrum, which causes the Cosmic Microwave Background (CMB) anisotropies and the large scale structure of the universe \cite{guth:1982}. Also, it predicts primordial gravitational waves. 

A lot of models of inflation have been proposed which are possible to classify by the number of free parameters \cite{martin:2014}. Each inflationary model is defined by a potential $V(\phi)$ that depends on a scalar field $\phi$. Those inflationary models have been tested with the observational data, and most of them have been ruled out by current observations \cite{escudero:2016}.

The Starobinsky  model \cite{starobinsky:1980} is the inflationary model that is currently supported by observations \cite{akrami:2018}. This model is a specific case of $n=2$ for inflation in $f(R)$ theories \cite{motohashi:2009,defelice:2010}.
Recently,  there has been interest in a particular class of power-law models that generalized the Starobinsky model \cite{motohashi:2015,chakravarty:2015,liu:2018,renzi:2020,canko:2020,cheong:2020,fomin:2020}, we call this the generalized Starobinsky model.  This model depends on a free parameter $p$, a real number close to unity, which can be tuned. For recovering the original form of the Starobinsky potential, we need to use $p=1$. Motohashi  \cite{motohashi:2015} has  found that the parameter $p$ is constrained to be $1.92 \lesssim 2 p \leq 2$.  Renzi \textit{et al.} \cite{renzi:2020} have constrained the cosmological parameters using the Planck data and have found that the parameter $p$ must be in the range  $0.962 \leq p \leq 1.016$.  In these references the analysis has been done using the slow-roll approximation.

The most used technique in  cosmological inflation is the slow-roll approximation, which consists that the potential $V(\phi)$ term will dominate over the kinetic term.  In this work, we use the slow-roll equations to set up the initial and final value of the scalar field, considering a number of e-folding of $N=60$. After that, we proceed to calculate numerically  the exact expression of  the scalar field $\phi$ and the scale factor $a$. Once obtained both expressions, they are used as initial conditions at $t=0$ in the numerical integration  of the perturbations equations.  Values of $p$ were taken between $0.995 \leq p \leq 1.005$, and  the cosmological parameters $A_\sca$, $n_\sca$ and $r$ were computed in all cases in order to  compare with the current observational data.

The article is structured as follows. In Section 2, we present the basic equations used in our calculations. In Section 3, we show the generalized Starobinsky model. In Section 4, we describe the method used to calculate the observables.  Section 5,  shows the results that we have obtained. Finally, in Section 6 we present the conclusions.

%*****************
\section{Basic equations}
%*****************

The equation of motion for an Universe dominated by a scalar field $\phi$ are given by:

\begin{eqnarray}
	\label{Friedmann}
	H^2&=&\dfrac{1}{3}\left[V(\phi)+\dfrac{1}{2}\dot{\phi}^2\right],\\
	\label{continuity}
	\ddot{\phi}&+&3H\dot{\phi}=-V,_\phi,
\end{eqnarray}
where dots  means derivative respect to the physical time $t$ and $V_{,\phi}$  derivative of the potential respect to the scalar field $\phi$.

\bigskip
Into the slow-roll approximation, Eqs. \eqref{Friedmann} and \eqref{continuity} reduce to

\begin{eqnarray}
	\label{sr2}
	H^2 &\simeq& \dfrac{1}{3} V(\phi),\\
	\label{sr1}
	3 H \dot{\phi} &\simeq& -V_{,\phi}.
\end{eqnarray}

The slow-roll parameter $\epsilon$ is given by 

\begin{equation}
	\label{epsilon}
	\epsilon=\dfrac{1}{2}\left( \dfrac{V_{,\phi}}{V}\right)^2,
\end{equation}
and the number of e-foldings between the horizon crossing of modes of interest and the end of inflation is giving by

\begin{equation}
	\label{N}
	N \simeq \int_{\phi}^{\phi_i} \dfrac{V}{V_{,\phi}} \D \phi.
\end{equation}

\bigskip
The scale factor $a$ and the scalar field $\phi$ exhibit a simpler form in the physical time $t$ than in the conformal time $\eta$, then we  write the equations for the scalar and tensor perturbations in variable $t$. The relation between $t$ and $\eta$ is given via the equation $\D t= a\D \eta$. Using this change of variable,  equations for scalar and  tensor perturbations can be written as \cite{truman:2020}

\begin{eqnarray}
	\label{dotuk}
	\ddot{u_k}&+&\dfrac{\dot{a}}{a}\dot{u_k}+\dfrac{1}{a^2}\left[k^2-\dfrac{\left(\dot{a}\dot{z_\sca}+a\ddot{z_\sca}\right)a}{z_\sca} \right]u_k=0,\\
	\label{dotvk}
	\ddot{v_k}&+&\dfrac{\dot{a}}{a}\dot{v_k}+\dfrac{1}{a^2}\left[k^2-\left(\dot{a}^2+a\ddot{a}\right) \right]v_k=0,
\end{eqnarray}
where $z_\sca=\sfrac{a \dot{\phi}}{H}$. Considering the limits  $k^2\gg|z_{S}''/z_{S}|$ (subhorizon scales) and $k^2\ll|z_{S}''/z_{S}|$  (superhorizon scales), we have  the  solutions to
Eq. (\ref{dotuk}) exhibit the following asymptotic behavior:

\begin{equation}
	\label{boundary_0}
	u_k\rightarrow \dfrac{e^{-ik\eta}}{\sqrt{2k}}
	\quad \left(k^2\gg|z_{S}''/z_{S}|, -k\eta\rightarrow \infty \right),
\end{equation}

\begin{equation}
	\label{boundary_i} u_k\rightarrow A_k z  \quad \left(k^2\ll|z_{S}''/z_{S}|,
	-k\eta\rightarrow 0\right).
\end{equation}

\noindent Equation \eqref{boundary_0} is used as the initial condition for the perturbations. The same asymptotic conditions hold for tensor
perturbations.

Once the solutions for $u_k$ and $v_k$ are known, the power spectra for scalar and tensor perturbations are given by the expressions

\begin{eqnarray}
	\label{PS}
	P_\sca(k)&=& \lim_{kt\rightarrow \infty} \dfrac{k^3}{2 \pi^2}\left|\dfrac{u_k(t)}{z_{S}(t)} \right|^2,\\
	\label{PT}
	P_\ten(k)&=& \lim_{kt\rightarrow \infty} \dfrac{k^3}{2 \pi^2}\left|\dfrac{v_k(t)}{a(t)} \right|^2,
\end{eqnarray}
these expressions are calculated at the superhorizon  $-k\eta \to 0$.

The spectral index for scalar perturbations is defined by:

\begin{equation}
	\label{nS}
	n_\sca(k)= 1+\dfrac{\D\ln P_\sca(k)}{\D\ln k}.
\end{equation}

\noindent In addition, the tensor-to-scalar ratio $r$ is defined as \cite{habib:2005b}

\begin{equation}
	\label{R}
	r=8\dfrac{P_T(k)}{P_S(k)}.
\end{equation}

%*****************
\section{The model}
%*****************

The generalized Starobinsky potential is given by \cite{martin:2014,renzi:2020}

\begin{equation}
	\label{gStarobinsky}
	V(\phi)= V_0 e^{-2 \sqrt{\frac{2}{3}}\phi} \left(e^{\sqrt{\frac{2}{3}}\phi}-1 \right)^{\frac{2p}{2p-1}},
\end{equation}
where $p$ is a real number close to unity, and

\begin{equation}
	V_0=6 \left(\dfrac{2p -1}{4p} \right) M^2 \left(\dfrac{1}{2p} \right)^{\dfrac{1}{2p-1}}.
\end{equation}

At $p=1$, equation \eqref{gStarobinsky} reduces to the Starobinsky potential \cite{martin:2014,mishra:2018}. The value of $M$ is fixed  in $M =1.30 \times 10^{-5}$  in order to obtain the parametrization of the amplitude for the scalar power spectrum at  the pivot scale $k=0.05$ Mpc$^{-1}$ in  the Starobinsky inflationary model $(p=1)$ \cite{canko:2020}.

\begin{figure}[th!]
	\centering
	\includegraphics[scale=0.40]{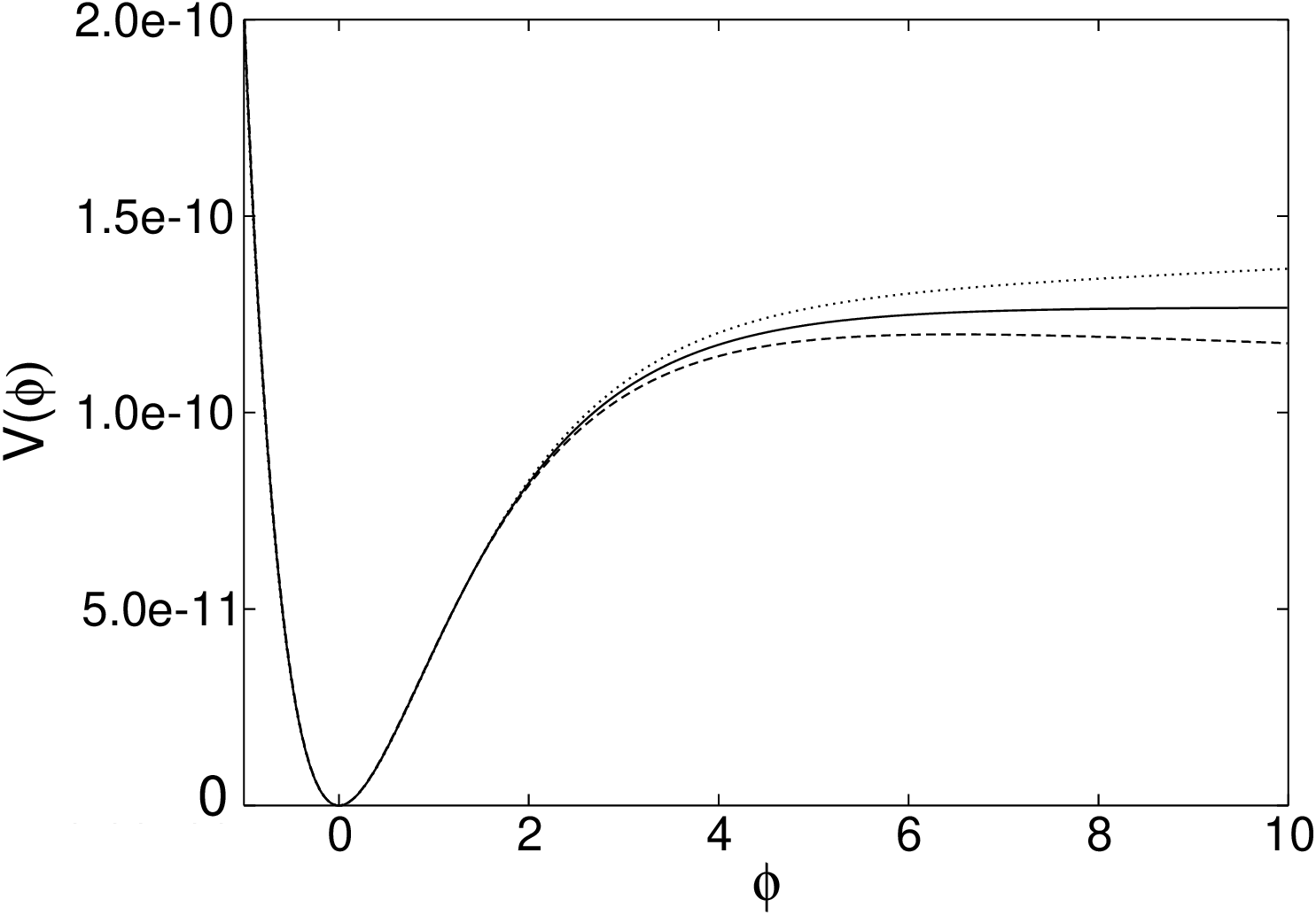}
	\caption{Generalized Starobinsky potential for $p=0.995$ (dotted-line), $p=1$ (Starobinsky potential, solid line), and $p=1.005$ (dashed line)}
	\label{potential}
\end{figure}

In Fig. \ref{potential} we show the form of the generalized Starobinsky potential for three different values of $p$.

%*****************
\section{Methods}
%*****************
We want to obtain the solution of Eqs. \eqref{Friedmann} and \eqref{continuity} to find the scalar field $\phi$ and the scale factor $a$. As initial conditions,  we use  the solution to the slow-roll equations \eqref{sr2} and \eqref{sr1}.

From Eq. \eqref{sr2} we obtain the scale factor into the slow-roll approximation  $a_\sr$ 

\begin{equation}
	\dfrac{\D a}{a}=\dfrac{1}{\sqrt{3}} \sqrt{V(t)} \, \D t\rightarrow  a_\sr=e^{h(t)},
\end{equation}
where

\begin{equation}
	h(t) = \dfrac{1}{\sqrt{3}} \int_{0}^{t} \sqrt{V(t)} \,\D t.
\end{equation}

From Eq. \eqref{sr1} we obtain the scalar field into the slow-roll approximation $\phi_\sr$,

\begin{equation}
	\label{dt}
    \D t = -\sqrt{3} \dfrac{\sqrt{V(\phi)}}{V_{,\phi}} \D \phi \rightarrow t =   g(\phi) \rightarrow g(\phi)-t=0,
\end{equation}
where

\begin{equation}
	\label{g}
    g(\phi)= -\sqrt{3}\int_{\phi_i}^{\phi}  \dfrac{\sqrt{V(\phi)}}{V_{,\phi}} \D \phi
\end{equation}

Doing the integration of Eq. \eqref{g} we obtain

\begin{eqnarray}
\nonumber
g(\phi)&=&-\dfrac{3}{M} \dfrac{(-1+2p)^{\sfrac{3}{2}}}{(-2+3p)} 2^{\frac{2-3p}{-1+2p}} p^\frac{1-p}{-1+2p} \\
\nonumber
          &\times& \left\{ \left(-1+e^{\sqrt{\frac{2}{3}}\phi}\right)^{\frac{2-3p}{1-2p}}\right.\\
\nonumber
          &\times& _{2}F_{1}\left[1,\dfrac{2-3p}{1-2p}, \dfrac{3-5p}{1-2p},\dfrac{(-1+p)}{p} \left(-1+e^{\sqrt{\frac{2}{3}}\phi} \right)\right]\\
\nonumber          
          &-& \left(-1+e^{\sqrt{\frac{2}{3}}\phi_i}\right)^{\frac{2-3p}{1-2p}}\\    
\nonumber                
          &\times& \left. _{2}F_{1}\left[1,\dfrac{2-3p}{1-2p}, \dfrac{3-5p}{1-2p},\dfrac{(-1+p)}{p} \left(-1+e^{\sqrt{\frac{2}{3}}\phi_i} \right)\right] \right\}.\\
\end{eqnarray}

 Using Mathematica$^\textnormal{\textregistered }$ we solve Eq. \eqref{dt} and obtain $\phi_\sr$ numerically, we call this :
 
$$
\phi_\sr(p,\phi_i,t).
$$

From Eq. \eqref{epsilon} we calculate the slow-roll parameter:

\begin{equation}
	\epsilon=\dfrac{4}{3} \dfrac{\left[1-2p+(p-1) e^{\sqrt{\dfrac{2}{3}}\phi}\right]}{(1-2p)^2 \left( e^{\sqrt{\dfrac{2}{3}}\phi}-1\right)},
\end{equation}
when $\epsilon=1$ we obtain the value of the scalar field at the end of inflation, $\phi_f$. Note that we recover the Starobinsky model when $p=1$.

From Eq. \eqref{N}, we obtain the number of e-folding $N$:

\begin{equation}
	\label{N_explicitly}
	N \simeq \dfrac{\sqrt{6}}{4} (\phi-\phi_i)-\dfrac{3}{4} \dfrac{p}{(p-1)} \ln \left[\dfrac{1-2p+(p-1)e^{\sqrt{\dfrac{2}{3}}\phi_i}}{1-2p+(p-1)e^{\sqrt{\dfrac{2}{3}}\phi}} \right].
\end{equation}

In the limit when $p=1$, Eq. \eqref{N_explicitly} reduces to

\begin{equation}
	\label{N_limit}
	N \simeq \dfrac{\sqrt{6}}{4} (\phi-\phi_i)-\dfrac{3}{4} \left(e^{\sqrt{\dfrac{2}{3}}\phi} -e^{\sqrt{\dfrac{2}{3}}\phi_i} \right).
\end{equation}

From Eq. \eqref{N_explicitly} and \eqref{N_limit} we can obtain numerically  the dependence  of $\phi$ with $N$, $\phi_\sr(N)$. Fixing $N=60$, we obtain the initial value of the scalar field $\phi_i$ for each value of $\phi_f$. The values obtained are shown in Table \ref{phi_table}.

\begin{table}[th!]
\begin{center}
\begin{tabular}{ccc}
\toprule
$p$                & $\phi_i$           &   $\phi_f$\\ 
\midrule
$0.995$        &$5.7246$          &$0.9473$\\      
$0.996$        &$5.6684$          &$0.9459$\\            
$0.997$        &$5.6132$          &$0.9444$\\            
$0.998$        &$5.5589$          &$0.9430$\\            
$0.999$        &$5.5056$          &$0.9416$\\            
$1.000$        &$5.4531$          &$0.9402$\\            
$1.001$        &$5.4016$          &$0.9388$\\            
$1.002$        &$5.3510$          &$0.9374$\\            
$1.003$        &$5.3013$          &$0.9360$\\            
$1.004$        &$5.2526$          &$0.9346$\\            
$1.005$        &$5.2046$          &$0.9332$\\                   
\bottomrule
\end{tabular}
\caption{Value of  $\phi_i$ and $\phi_f$ to several values of $p$.}
\label{phi_table}
\end{center}
\end{table}

\begin{figure}[th!]
\centering
\includegraphics[scale=0.40]{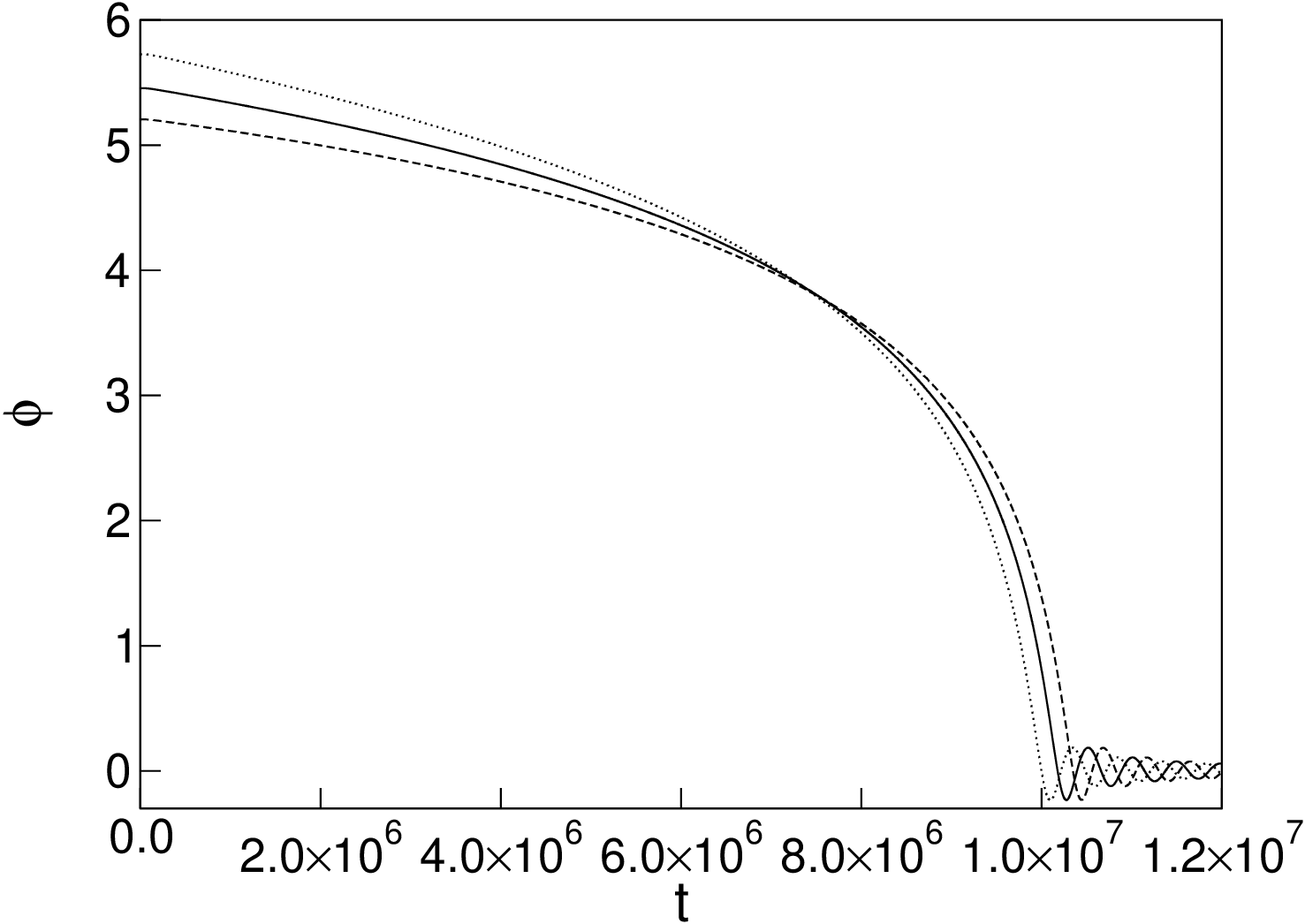}
\caption{Scalar field $\phi$ in function of the cosmic time $t$ for $p=0.995$ (dotted-line), $p=1$ (solid line), and $p=1.005$ (dashed line)}
\label{phi_ex}
\end{figure}

\begin{figure}[th!]
\vspace{-0.5cm}
\centering
\includegraphics[scale=0.40]{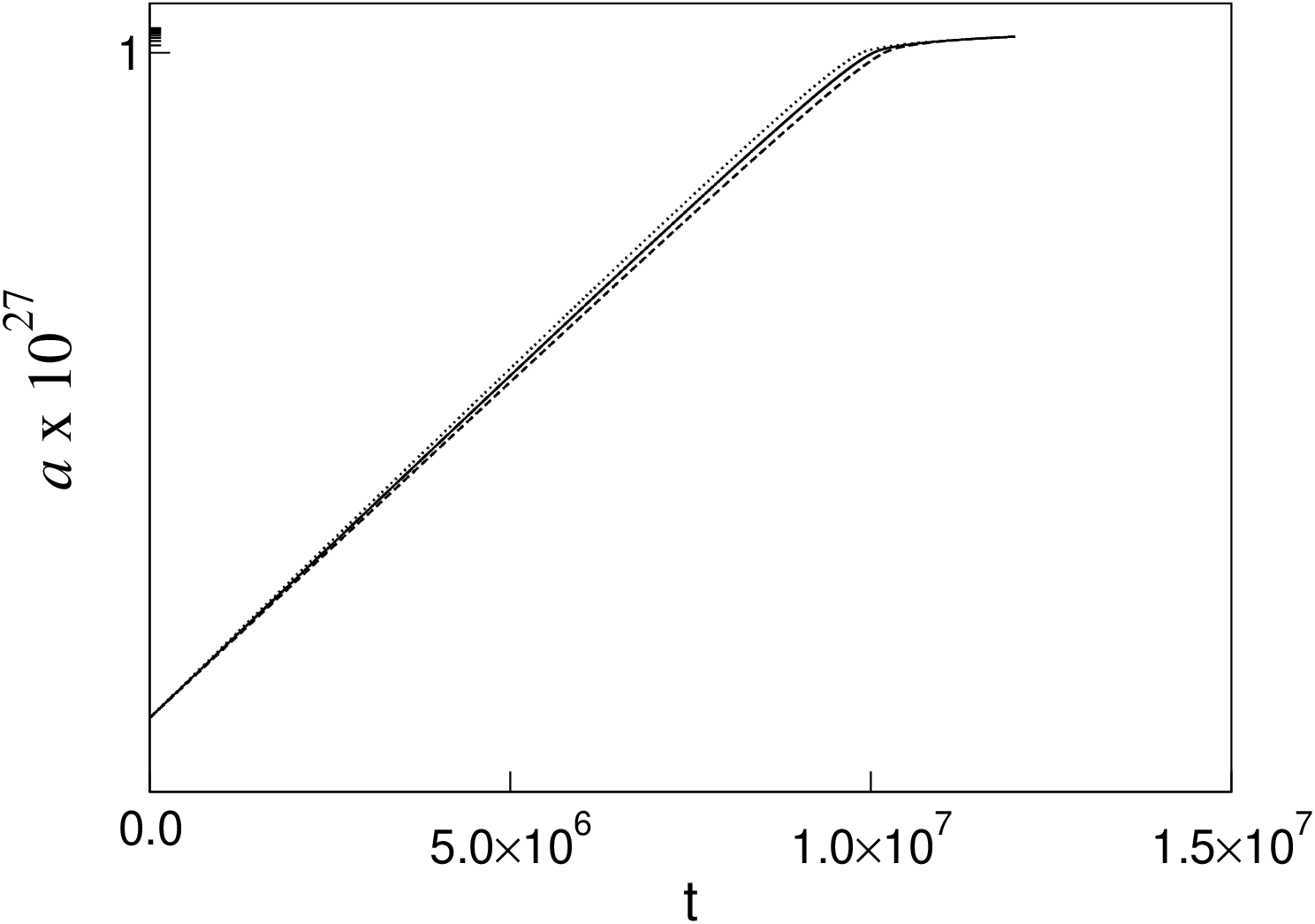}
\caption{Scale factor $a$ in function of the cosmic time $t$ for $p=0.995$ (dotted-line), $p=1$ (solid line), and $p=1.005$ (dashed line)}
\label{a_ex}
\end{figure}

The values of $\phi_\sr(p,\phi_i,0)$ and $\dot{\phi}_\sr(p,\phi_i,0)$  are used in the numerical code as initial conditions to solve the  system of equations \eqref{Friedmann} and \eqref{continuity} in order to find the exact function of the scale factor  $a_\ex(t)$  and the exact function of the scalar field $\phi_\ex(t)$ in function of the cosmic time $t$. Figs. \ref{phi_ex} and \ref{a_ex}  show the scale factor and the scalar field in function $t$. When the scalar field begins to oscillate the inflation ends. Note that in this model the expansion is quasi exponential, unlike  the de Sitter Universe.

The functions $a_\ex(t)$  and $\phi_\ex(t)$ are used to do the numerical integration of equations \eqref{dotuk} and \eqref{dotvk},   and then using Eqs. \eqref{PS} and \eqref{PT} we obtain $P_\sca(k)$ and $P_\ten(k)$  \cite{truman:2020}. By using the value of the scalar power spectrum, we calculate  from Eq. \eqref{nS} the scalar spectral index at the pivot scale of $k=0.05$ Mpc$^{-1}$. With the help of the tensor power spectrum using Eq. \eqref{R}, we calculate the scalar-to-tensor ratio at the pivot scale of $k=0.002$ Mpc$^{-1}$. In the following we call $P_\sca$ as $A_\sca$.
%*****************
\section{Results and discussion}
%*****************

We do slight variations of the parameter $p$ around $p=1$, from $p=0.995$ to $p=1.005$. In Table \ref{p_1_table} we show the results for  $p=1$, note that the values of the cosmological parameters are in good agreement  with those reported in the literature when $p=1$ \cite{renzi:2020}.

\begin{table}[th!]
\begin{center}
\begin{tabular}{cc}
\toprule
Parameter                                           & Value\\ 
\midrule 
$A_\sca$                                           & $\;\;2.1442 \times 10^{-9}$              \\
$\ln\left(10^{10} A_\sca \right) $   & $\;\;3.065$   \\
$n_\sca$                                            & $\;\;0.9638$ \\
 $r_{0.002} $                                     & $\;\;0.00345$  \\
\bottomrule
\end{tabular}
\caption{Cosmological parameters for the Starobinsky inflationary model $(p=1)$. The value of  $n_\sca$ is calculated at the pivot scale $k=0.05$ Mpc$^{-1}$.}
\label{p_1_table}
\end{center}
\end{table}

We have found  that the value of $p$ that reproduces the value supported by the current observational data \cite{renzi:2020}  is $p=1.0004$. The results are shown  in Table \ref{p_1.0004_table}, note that the values of $A_\sca$,  $n_\sca$ and $r$ are in agreement with the current observational data \cite{renzi:2020} .

\begin{table}[htbp]
\begin{center}
\begin{tabular}{cc}
\toprule
Parameter                                           & Value\\ 
\midrule 
$A_\sca$                                           & $\;\;2.2023 \times 10^{-9}$              \\
$\ln\left(10^{10} A_\sca \right) $   & $\;\;3.092$   \\
$n_\sca$                                            & $\;\;0.9632$ \\
$r_{0.002} $                                     & $\;\;0.00335$  \\
\bottomrule
\end{tabular}
\caption{Cosmological parameters for the  generalized Starobinsky inflationary model for $p=1.0004$. The value of  $n_\sca$ is calculated at the pivot scale $k=0.05$ Mpc$^{-1}$.}
\label{p_1.0004_table}
\end{center}
\end{table}

\begin{figure}[th!]
\centering
\includegraphics[scale=0.40]{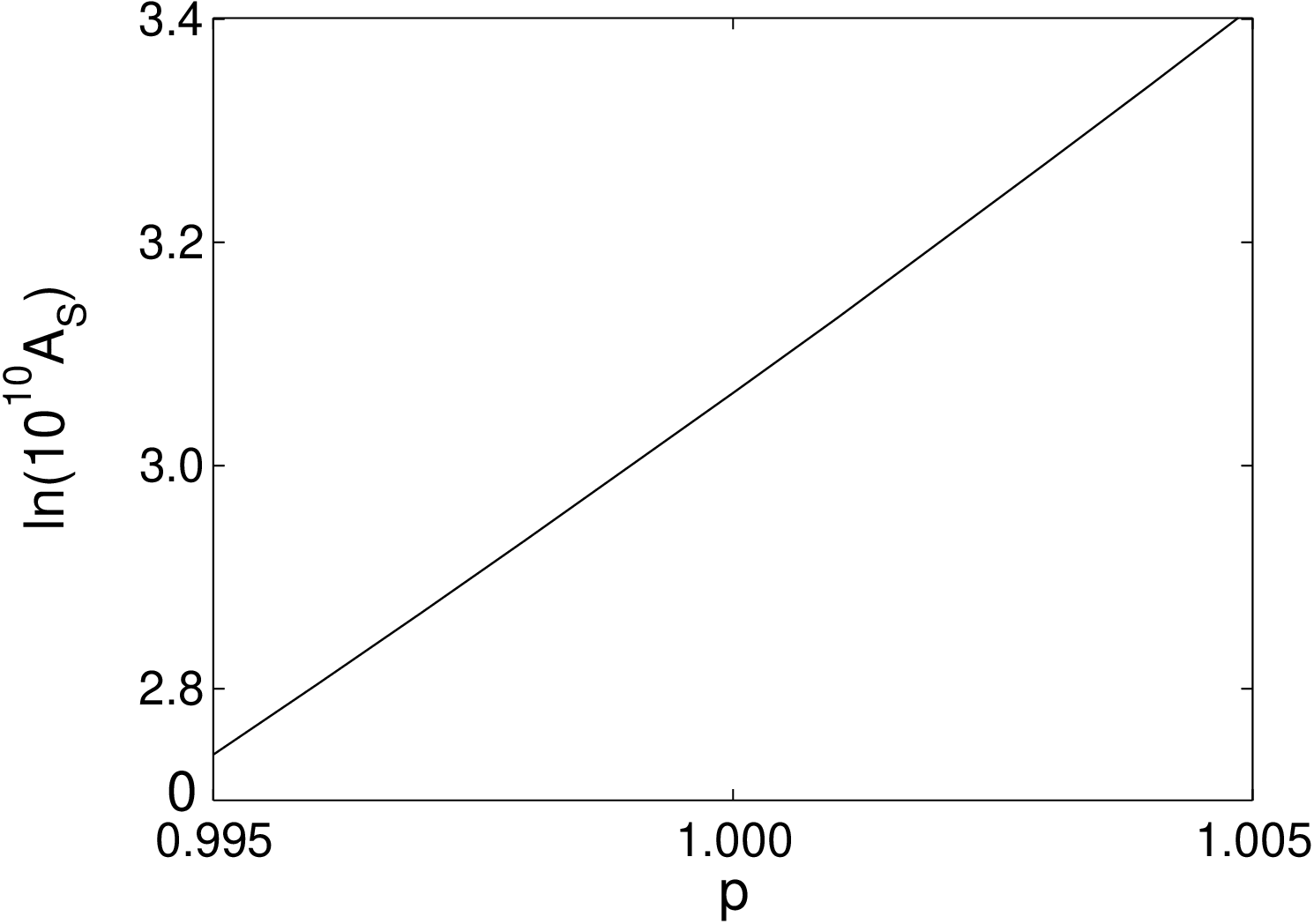}
\caption{$\ln\left(10^{10} A_\sca\right)$ for $k=0.05$ Mpc$^{-1}$  in function of the parameter $p$.}
\label{fig:AS}
\end{figure}

\begin{figure}[th!]
\centering
\includegraphics[scale=0.40]{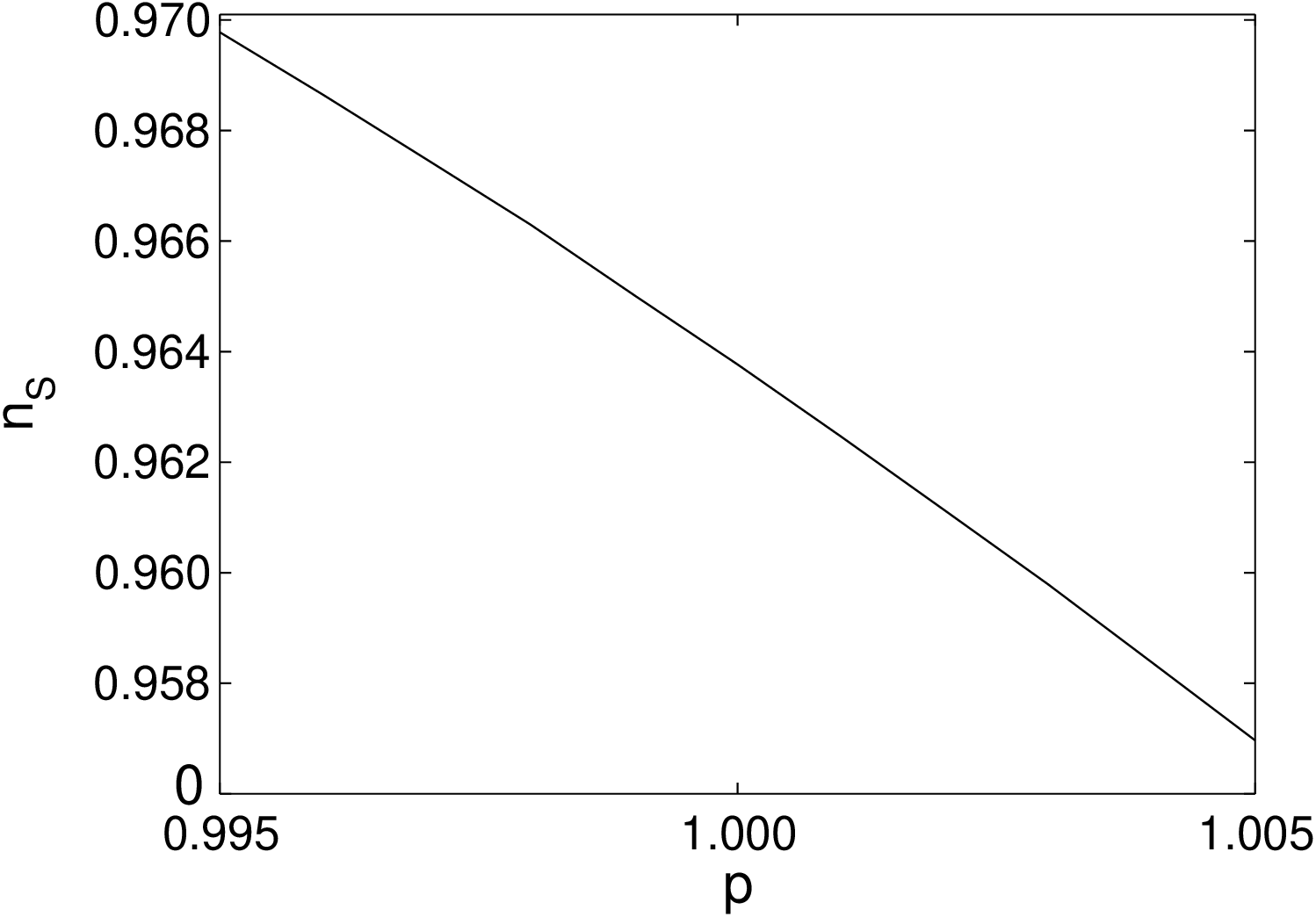}
\caption{Scalar spectral index $n_\sca$ for $k=0.05$ Mpc$^{-1}$  in function of the parameter $p$.}
\label{fig:nS}
\end{figure}

\begin{figure}[th]
\centering
\includegraphics[scale=0.40]{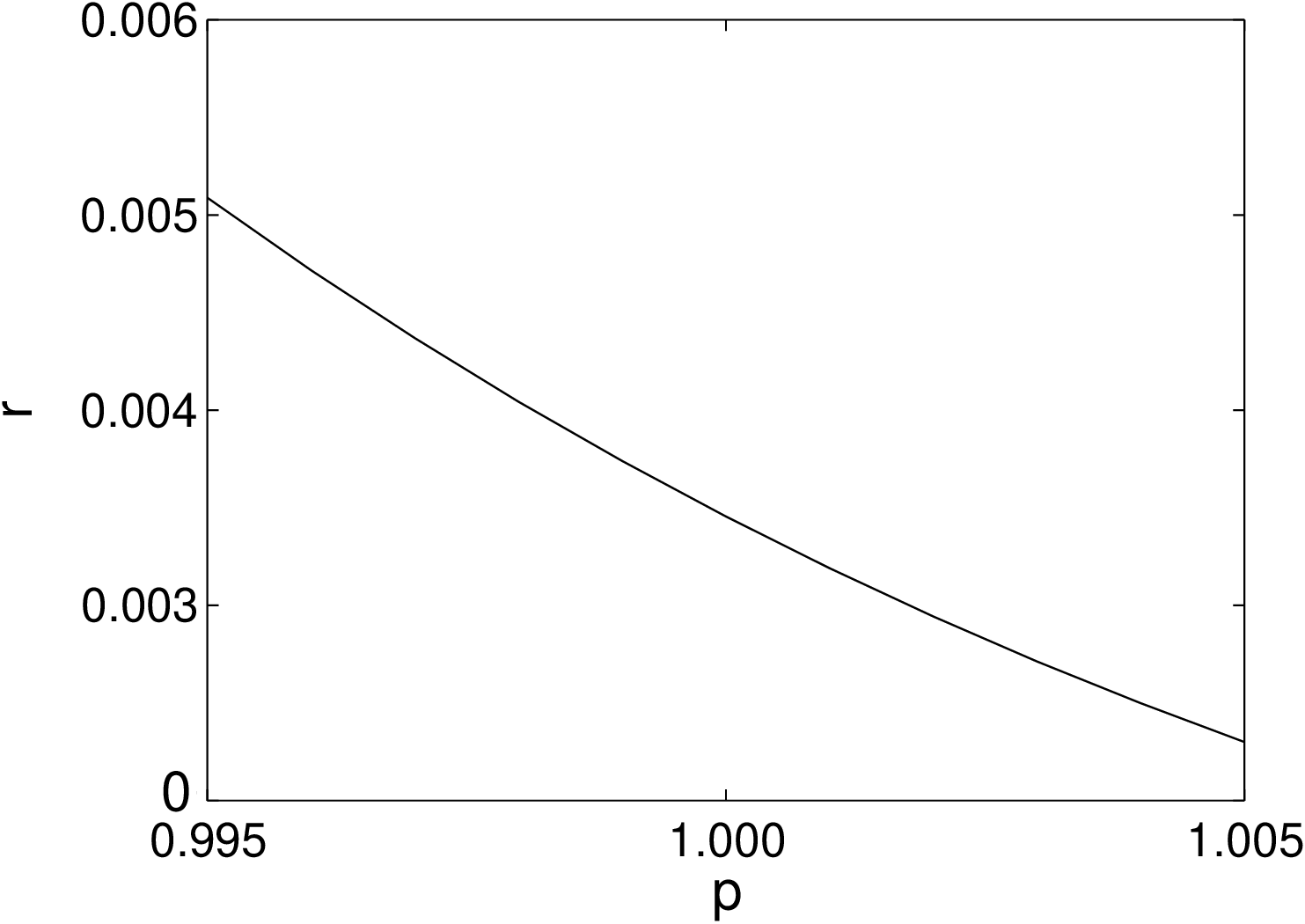}
\caption{Scalar-tensor ratio $r$ for $k=0.002 $Mpc$^{-1}$ in function of the parameter $p$.}
\label{fig:r}
\end{figure}

In Fig. \ref{fig:AS} we can observe the dependence of $\ln\left(10^{10} A_\sca\right)$ for $k=0.05$ Mpc$^{-1}$  with the parameter $p$, which increases as $p$ increases. Fig. \ref{fig:nS} shows the behavior of $n_\sca$ for $k=0.05$ Mpc$^{-1}$  in function of the parameter $p$, and we can observe  that $n_\sca$  decreases as $p$ increases.
In Fig. \ref{fig:r}, we present the scalar-tensor ratio $r$ for $k=0.002 $ Mpc$^{-1}$  in function of the parameter $p$. This value decreases when $p$ increases.

\begin{figure}[th]
\centering
\includegraphics[scale=0.40]{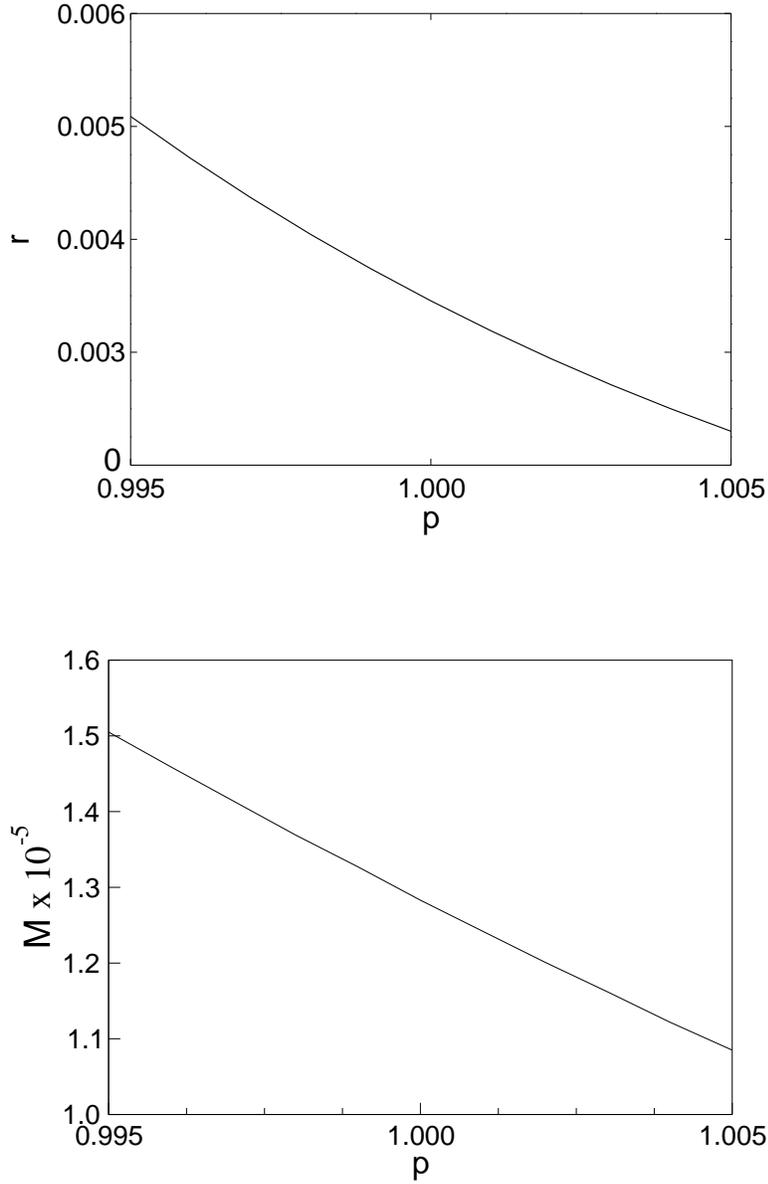}
\caption{Value of $M$ in function of the parameter $p$.}
\label{M}
\end{figure}

As an additional analysis, we adjusted the energy scale $M$ of the inflationary model for each value of $p$ in order to satisfy the observational constraint on $A_S$. In Fig. \ref{M}  we can observe that for small values of $p$ it is necessary to increase the value of $M$ for obtaining an accuracy with the observational data. For large values of $p$, it is needed to decrease the value of $M$ for matching with reported data.

%*****************
\section{Conclusions}
%*****************

In this work, we have studied a particular class of power-law models that generalize the Starobinsky model through a parameter $p$. We made a sweep over parameter $p$ between $0.999 <  p  <1.0005$ and found that the value of $p= 1.0004$ reproduces the value of the cosmological parameters according to observations. On the other hand, when we fit the energy scaling $M$ for each value of $p$ we obtain that the value of $M$ decreases as $p$ increases.

%*****************

%*****************
\section{Acknolegment}
%*****************

The authors want to express their gratitude to Fabrizio Renzi and Werner Br\"amer-Escamilla for useful discussions. 
%*****************

\bibliographystyle{unsrt}
%*****************
%\bibliography{gStarobinsky}

\begin{thebibliography}{10}
	
	\bibitem{liddle:2000}
	A.~R. Liddle and D.~H. Lyth.
	\newblock {\em {Cosmological inflation and large-scale structure}}.
	\newblock Cambridge University Press, 2000.
	
	\bibitem{guth:1981}
	A.~H. Guth.
	\newblock {Inflationary universe: A possible solution to the horizon and
		flatness problems}.
	\newblock {\em Phys. Rev. D}, 23:347, 1981.
	
	\bibitem{martin:2019}
	J.~Martin.
	\newblock {Cosmic Inflation: Trick or Treat?}
	\newblock {\em arXiv:1902.05286}, 2019.
	
	\bibitem{guth:1982}
	{A. H. Guth and So-Young Pi}.
	\newblock {Fluctuations in the new inflationary universe}.
	\newblock {\em Phys. Rev. Lett.}, 49:1110, 1983.
	
	\bibitem{martin:2014}
	{J. Martin, C. Ringeval, and V. Vennin}.
	\newblock {Encyclopaedia Inflationaris}.
	\newblock {\em Phys. Dark Univ.}, 5-6:75--235, 2014.
	
	\bibitem{escudero:2016}
	{M. Escudero, H. Ramírez, L. Boubekeur, E. Giusarma, and O. Mena}.
	\newblock {The present and future of the most favoured inflationary models
		after Planck 2015}.
	\newblock {\em JCAP}, 02:020, 2016.
	
	\bibitem{starobinsky:1980}
	A.~A. Starobinsky.
	\newblock {A new type of isotropic cosmological models without singularity}.
	\newblock {\em Phys. Lett. B}, 91:99, 1980.
	
	\bibitem{akrami:2018}
	Y.~Akrami \textit{et al.}
	\newblock {Planck 2018 results. X. Constraints on inflation}.
	\newblock {\em arXiv:1807.06211}, 2018.
	
	\bibitem{motohashi:2009}
	{H. Motohashi, A. A. Starobinsky, and J. Yokoyama}.
	\newblock {Analytic solution for matter density perturbations in a class of
		viable Cosmological $f(R)$ models}.
	\newblock {\em Int. J. Mod. Phys. D}, 18:1731, 2009.
	
	\bibitem{defelice:2010}
	{A. De Felice and S. Tsujikawa}.
	\newblock {f(R) Theories}.
	\newblock {\em Living Rev. Relativity}, 13:3, 2010.
	
	\bibitem{motohashi:2015}
	{H. Motohashi}.
	\newblock {Consistency relation for $R^p$ inflation}.
	\newblock {\em Phy. Rev. D}, 91:064016, 2015.
	
	\bibitem{chakravarty:2015}
	{G. K. Chakravarty and S. Mohanty}.
	\newblock {Power law Starobinsky model of inflation from no-scale SUGRA}.
	\newblock {\em Phys. Lett. B}, 746:242, 2015.
	
	\bibitem{liu:2018}
	{Lei-Hua Liu}.
	\newblock {Analysis of $R^p$ inflationary model as $p \geq 2$}.
	\newblock {\em arXiv:1807.00666v3}, 2018.
	
	\bibitem{renzi:2020}
	{F. Renzi, M. Shokri, and A. Melchiorri}.
	\newblock {What is the amplitude of the gravitational waves background expected
		in the Starobinsky model?}
	\newblock {\em Phys. Dark. Univ.}, 27:100450, 2020.
	
	\bibitem{canko:2020}
	{D. D. Canko, Ioannis D. Gialamas, and G. P. Kodaxis}.
	\newblock {A simple $F(\mathcal{R},\phi)$ deformation of Starobinsky
		inflationary model}.
	\newblock {\em Eur. Phys. J. C.}, 80:458, 2020.
	
	\bibitem{cheong:2020}
	{D. Y. Cheong, H. M. Lee and S. C. Park}.
	\newblock {Beyond the Starobinsky model for inflation}.
	\newblock {\em Phys. Lett. B}, 805:135453, 2020.
	
	\bibitem{fomin:2020}
	{I. V. Fomin, S, V, Chervon, and A. V, Tsyganov}.
	\newblock {Generalized scalar-tensor theroy of gravity reconstruction from
		physical potentiasl of a scalar field}.
	\newblock {\em Eur. Phys. J. C.}, 80:350, 2020.
	
	\bibitem{truman:2020}
	{T. Tapia, M. Z. Mughal, and C. Rojas}.
	\newblock {Semiclassical analysis of the Starobinsky inflationary model}.
	\newblock {\em Phys. Dark Univ.}, 30:100650, 2020.
	
	\bibitem{habib:2005b}
	{S. Habib and A. Heinen and K. Heitmann and G. Jungman}.
	\newblock {Inflationary Perturbations and Precision Cosmology}.
	\newblock {\em Phys. Rev. D}, 71:043518, 2005.
	
	\bibitem{mishra:2018}
	{S. S. Mishra, V. Sahni, and A. V. Toporensky}.
	\newblock {Initial conditions for inflation in an FRW universe}.
	\newblock {\em Phys. Rev. D}, 98:083538, 2018.
	
\end{thebibliography}

\end{document}